\documentclass[12pt]{article}

\def\p{\partial}
\def\rar{\rightarrow}
\usepackage{graphicx,calrsfs,color,dutchcal,eufrak}
\usepackage{amssymb,amsmath,bm,amsthm,todonotes}
\newtheorem{theorem}{Theorem}

\def\vb{{\bf v}}
\def\u{{\bf u}}
\def\k{{\bf k}}
\def\bp{{\bf p}}

\def\V{{\bf V}}
\def\bU{{\bf U}}
\def\X{{\bf X}}
\def\Y{{\bf Y}}
\def\Z{{\bf Z}}

\def\d{{\mathsf{div}}}
\def\bR{{\mathbb{R}}}
\def\bC{{\mathbb{C}}}
\def\bZ{{\mathbb{Z}}}

\def\ups{\upsilon}
\def\U{{\cal U}}
\def\W{{\cal W}}

\def\x{{\bf x}}
\def\y{{\bf y}}
\def\z{{\bf z}}

\def\l{{\mathcal l}}

\def\L{{\mathcal L}}
\def\H{{\mathcal H}}
\def\bea{\begin{eqnarray}}
\def\eea{\end{eqnarray}}
\def\be{\begin{equation}}
\def\ee{\end{equation}}

\newsavebox{\astrutbox}
\sbox{\astrutbox}{\rule[-5pt]{0pt}{20pt}}

\newtheorem{remark}{Remark}
\newtheorem{example}{Example}

\newtheorem*{postulates*}{Postulates}
\begin{document}

\title{A Transfer Operator Approach to Relativistic Quantum Wavefunction}

\author{Igor Mezi\'c \\
University of California, Santa Barbara, CA 93106-5070, USA\\
mezic@ucsb.edu}
\vspace{10pt}

\maketitle

\begin{abstract}
The original intent of the Koopman-von Neumann formalism
was to put classical and quantum mechanics on the same footing by introducing an operator formalism into
classical mechanics. Here we pursue their path the opposite way and examine what  transfer 
operators can say about quantum mechanical evolution.
To that end, we  introduce a physically motivated scalar wavefunction formalism for a velocity field on a 4-dimensional pseudo-Riemannian manifold, and obtain an evolution equation for the associated wavefunction, a generator for an associated weighted transfer operator. The generator of the scalar evolution is of first order in space and time.  The probability interpretation of the formalism leads to recovery of the Schr\"odinger equation  in the non-relativistic limit. In the special relativity limit, we show that the scalar wavefunction of  Dirac spinors satisfies the new equation.  A connection with string theoretic considerations for mass is provided. \end{abstract}
\tableofcontents
%\maketitle

\section{Introduction}
Dynamical systems theory can be pursued in the phase-space (Poincar\'e) formalism \cite{guckenheimer1984nonlinear}, or alternatively in the  Koopman formalism \cite{Koopman:1931,Mezic:2005,Budisicetal:2012}. The Koopman formalism applied in phase space leads to the probability interpretation of the associated phase-space wavefunction  consistent with the Born interpretation in quantum mechanics \cite{wilczek:2015}. Born's proposal on interpretation of the square of the wavefunction as probability led to successful application of quantum mechanics to a broad swath of problems. The dichotomy between the phase-space domain of the classical wavefunction and the physical spacetime nature of the quantum wavefunction recently led to a number of efforts to reconcile the two (see e.g. \cite{antoniou2002implementability,gozzi2002minimal,wilczek:2015,ghose:2017,klein2018koopman,bondaretal:2019,viennot2018schrodinger,joseph2020koopman,mauro2002koopman} and the rest of the articles in this volume). These works are pursued in the nonrelativistic context. A different approach was pursued in \cite{giannakis2019quantum}, where the spectrum of the quantum harmonic oscillator was related to the Koopman operator spectrum of the classical harmonic oscillator by a construction involving a pair of harmonic oscillators with Hamiltonians of opposite sign.

In this paper we pursue the operator-theoretic approach  to derive  an equation of motion - the relativistic quantum transfer equation (RQTE) -  for the resulting quantum-theoretical wavefunction starting from a relativistic dynamical system on a 4-dimensional space-time. Namely, we start from the spacetime manifold, and not the phase space, and utilize Fock's proper-time formalism \cite{fock1937proper}. The key idea is that the RQTE arises from the projection of a 4-dimensional conserved field  through a complex scalar field. 
The resulting equation - when presented in the probabilistic interpretation - has solutions that reduce to the Schr\"odinger equation in the nonrelativistic limit, and the euation for  the Dirac scalar in the special relativity limit. 

 The paper is organized as follows: in section \ref{sec:prel} we introduce the relativistic setting and the notation. In section \ref{sec:scalarwave} we derive RQTE under several postulates. We describe the class of operators - the weighted composition operators - that are generated by RQTE.  In section \ref{sec:Dirac} we discuss the relationship between RQTE and the Dirac equation. In section \ref{sec:exa} we consider several examples treated within the RQTE formalism: harmonic oscillator, particle in a box and Gaussian wavepacket. We discuss the relationship of the RQTE wavefunction with mass in Appendix \ref{sec:mass} and relationship with notion of mass in string theory in Appendix \ref{sec:string}.
\section{Preliminaries} 
 \label{sec:prel}
Let $M$ denote a $4$-dimensional space-time pseudo-Riemannian manifold endowed with a metric tensor $g$. Consider the section of its tangent bundle $TM$, the  proper velocity field (the four-velocity field) $\V=d\X/d\tau$ \cite{dirac1996general} where $\X(\tau):\bR\rar M$ is the time-like world line parametrized by the proper time $\tau$. 
We define the level sets of proper time $\tau$ on $M$  to be able to use it for  evolution of the flow of $\V$. Any vector field on $M$ can be rectified near a point $\X$ with $\V(\X)\neq 0$ \cite{olver2000applications}. Since the four-velocity field $\V$ is nonzero everywhere, there exists a neighborhood $\mathcal{N}_\X$ of any point $\X$ in which it can be rectified by a local choice of coordinates $(x_0(\X),...,x_3(\X))$ on $M$. In the coordinates $(x_0,...,x_3)$, the four velocity field has components $\V=(c,0,0,0)$. Note that $(x_1,x_2,x_3)$ label points on the $x_0=0$ intersection of an individual world line. Let $\sigma=\tau(0,x_1,x_2,x_3)$ be the proper time field over the section $x_0=0$. We can define a new parameter $s=\tau-\sigma$ in a small neighborhood of $\X$. In this way, the $0$ proper time is synchronized for all trajectories in a neighborhood. Absent topological obstructions, this can be extended to the whole of $M$ to define a space slice $M^\tau_S$. With topological obstructions, the construction is still valid on subsets of $M$. In this case, we redefine $M$ to be such a subset. We keep the notation $\tau$ for the reparametrized proper time.   The norm of $\V$ defined using the metric tensor $g$ on $M$ is constant,  $||\V||^2=-c^2,$ where $c$ is the speed of light in vacuum \cite{fock1937proper} (we are using the $(-1,1,1,1)$ metric convention). We denote by $G^\tau:M\rar M$  the flow of   $\V$ on $M$. We denote by  $D_\tau f$  the proper time derivative (i.e. the Lie derivative \cite{olver2000applications}) of $f$, representing the change of a scalar physical quantity in the direction of $\V$. The manifold is equipped  with the volume form with density $\sqrt{|\det g|}$.

The flow $G^\tau$ can be used to define the family of Koopman composition operators \cite{Koopman:1931} parametrized by $\tau$  acting on (in general, complex) functions $f:M\rar \bC$ by
\be
{\U}^\tau f(\X)=f\circ G^\tau(\X).
\ee
Note that, in contrast with Koopman's original formulation on the phase space, $\U^\tau$ acts on functions defined on the spacetime $M$.
The operator $D_\tau$ is the generator of the evolution $\U^\tau$. The functions in the eigenspace at $0$ of $D_\tau$ are conserved quantities, since 
\be
D_\tau f=0
\ee
implies $f$ is conserved on the world line $\X(\tau)$. In terms of the Koopman operator evolution, for such $f$ we get
\be
{\U}^\tau f(\X)=f(\X).
\ee
In line with the terminology used in Koopman operator theory \cite{MezicandBanaszuk:2004,Mezic:2005} we call functions $g:M\rar \bC$ {\it observables}. By identification with the associated, position-dependent operators, the terminology is consistent with that of quantum mechanics.  

\section{Wavefunction Evolution}
\label{sec:scalarwave}
\begin{figure}[h!]
\centering
\includegraphics[clip=true, trim=100 90 200 90,
height=2in, width=4in ]{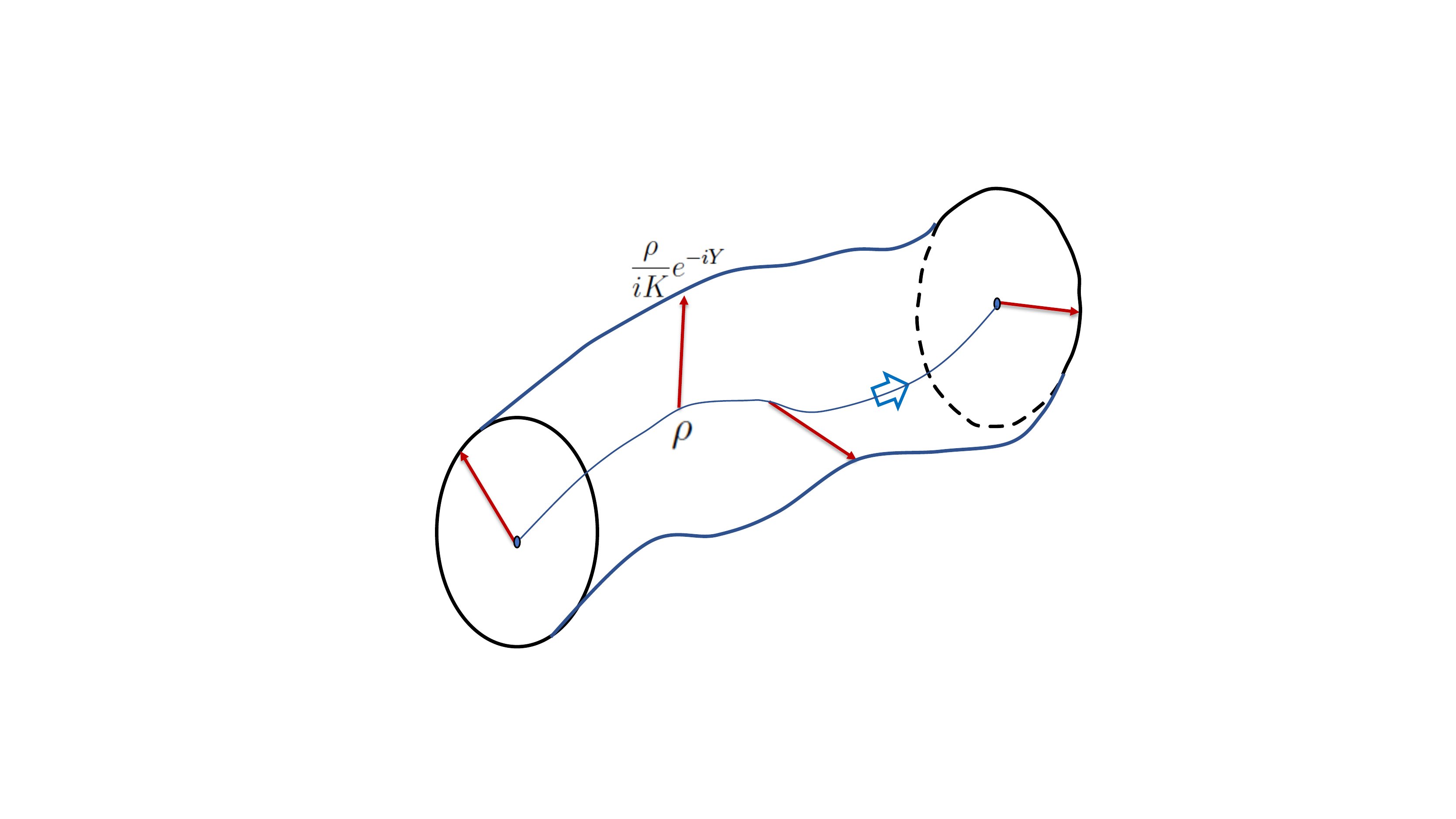}
\caption{The geometry of the fiber bundle over $M$. } \label{fig:fiber}
\end{figure}

Consider a  field $\rho$ conserved under trajectories of $\V$ on $M$. Its  restriction  onto level sets of the complex field  $e^{iY}$  of modulus $1,$  with phase $Y$ reads
\be
\frac{\rho}{De^{iY}}=\frac{\rho}{i|DY|e^{iY}}.
\label{eq:rest}
\ee
We assume that the density $\rho$ is
not observed directly, but is projected via a complex scalar field $e^{iY}$, as indicated by equation (\ref{eq:rest}) and
shown in figure \ref{fig:fiber}. The geometry can be described as that of a fiber bundle over $M$ and $\rho e^{iY}$ is a horizontal lift of the spacetime trajectory. This construction renders the appearence of complex numbers in quantum mechanics a natural consequence of geometry.

Given this geometric formulation, we use the following postulates:
\vskip .3cm
\subsection{Postulates}
\begin{enumerate}
\item \label{post1} There is a function $\rho:M\rar \bR$   that is constant on trajectories of $\V$ satisfying
\be
D_\tau \rho=0. 
\label{eq:rho1}
\ee
We argue in the  Appendix \ref{sec:mass} that $\rho$ is physically the oscillation wavenumber and is related to mass (and thus energy).
\item  \label{post2} The observable wavefunction $\psi$ is the pushforward of  $\rho$ by an observable $e^{iY}$ given by
\be
 \psi=\frac{\rho}{i|DY|e^{iY}}=\frac{\rho}{iKe^{iY}}=\frac{\rho }{iK}e^{-iY}.
  \label{eq:OW1}
 \ee
 where $Y$ is a phase and $K=|DY|$. 
 This, in turn, implies
 \be
 \rho=iKe^{iY}\psi.
 \ee
 \item $\rho/|DY|$ is an invariant density for $\V$.
\end{enumerate}
\begin{remark}The last postulate is natural in view of the fact that - when extending the classical action - velocity relationship relativistic, and identifying $Y$ with relativistic action, $DY$ is proportional to the space-time velocity $\V$. Thus, the density   $\rho/|DY|$ is inversely proportional to the velocity magnitude and thus is invariant.
 \end{remark}
%\begin{remark}  \label{post2} In the Born rule, the  wavefunction modulus $\varphi\varphi^c$ is the pushforward of  $\sqrt\rho$ by a real observable phase $Y$ given by
%\be
% \varphi\varphi^c=\frac{\rho}{|DY|}.
%  \label{eq:OW1}
% \ee
% and thus the wavefunction $\varphi$ satisfies
%\be
% \varphi=\frac{\sqrt\rho}{i|DY|^{1/2}e^{iY}}=\frac{\sqrt\rho}{i\sqrt{K}e^{iY}}.
%  \label{eq:OW1}
% \ee
% where $Y$ is a phase and $K=|DY|$. 
% This, in turn, implies
% \be
% \rho^{1/2}=i\sqrt{K}e^{iY}\varphi.
% \ee
%Indicating the relationship between our ``observable wavefunction" $\psi$ and the quantum mechanical wavefunction as.
%\be
%\frac{\psi}{\varphi}=\sqrt{\frac{\rho}{K}}
%\ee
%As a consequence of these relationships, the standard question in other ``mechanistic" theories of quantum mechanics such as deBroglie-Bohm theory,
%where the initial relationship between probability and pilot wavefunction $p=|\psi|^2$ is postulated as ``quantum equilibrium" hypothesis \cite{bohm1982broglie}  is obtained here as a simple
%consequence of the fact that we are treating $\rho$ as the physical oscillation density, that is observed with higher probability if the oscillation density is higher, and the normalized version of $\varphi$ is used to construct the probability that satisfies the ``quantum equilibrium".
%\end{remark}

Note that the factor $i$ is used in the wavefunction definition just for convenience of the calculations below since 
the constant phase of the wavefunction is irrelevant. 
 
Under the above assumptions, we have
\begin{theorem} Let   $Y=-S/\hbar, $ where $\hbar$ is the reduced Planck constant, and  $D_\tau S=\L$, analogous to the standard notions of the action $S$ and the Lagrangian $\L$. The wavefunction $\psi$ satisfies
\be
i\hbar D_\tau \psi=-\L\psi - i\hbar{\mathsf{div}} \V\psi
\label{eq:all}
\ee
where, in coordinates,
\be
\d \V=\frac{1}{\sqrt|g|}\sum_j\frac{\p \sqrt|g|V^j}{\p X_j}
\ee
is divergence with respect to volume element, where $|g|=|\det g_{ij}|$ is the absolute value of the determinant of the  metric tensor.

\end{theorem}
\begin{proof}
By assumption 1. $D_\tau \rho=0$, and we have
\bea
i\hbar D_\tau \psi&=&-\hbar\rho\frac{D_\tau K}{K^2} e^{-iY} -i\hbar\frac{\rho  }{K} D_\tau Y  e^{-iY} \nonumber \\
                           &=&-i\hbar\frac{D_\tau K}{K} \psi +\hbar D_\tau Y  \psi  \nonumber \\
                           &=&-i\hbar\frac{D_\tau K}{K} \psi -\L  \psi.
                           \label{eq:spt1}
\eea
Now we show that $K=|DY|$ must satisfy
\be 
D_\tau |DY|=|DY|\d \V.
\label{eq:proj}
\ee
Since $\rho/|DY|$ is an invariant density, 
\be
D_\tau (\rho/|DY|)=-\rho|DY|^{-2}D_\tau |DY|=-\rho/|DY|\d\V
\ee
implying
\be
D_\tau |DY|=|DY|\d\V.
\label{dy}
\ee

Now (\ref{eq:spt1}) yields
\be
i\hbar D_\tau \psi=[-\L-i\hbar\  {\mathsf{div}} \V]\psi.
\label{eq:all1}
\ee
where $D_\tau$ is the proper time derivative, $\L$ is the Lagrangian, and ${\mathsf{div}}$ is the  divergence of the  vector field  $\V$ with respect to $\sqrt|g|$.
\end{proof}

It is notable that (\ref{eq:all}) has the solution
\bea
\psi(\Y,\tau)&=&\psi_0(G^{-\tau}(\Y))e^{-\int_0^\tau {\mathsf{div} \V(G^{s}(G^{-\ups}(\Y)))}ds}e^{iS(\tau,\Y)/\hbar}, \nonumber \\
&=&\psi_0(\X_0)e^{-\int_0^\tau {\mathsf{div} \V(G^{s}(\X_0))}ds}e^{iS(\tau,\Y)/\hbar} ,
\label{eq:Feyn}
\eea
where $\psi(\Z,0)=\psi_0(\Z),$ and $\X_0=\X(G^{-\tau}(\Y))$ is the initial position at $\tau=0$ of trajectory landing at $\Y$ at $\tau$.
\subsection{Relationship with the Schr\"odinger Equation} 
Note that for any power $(\rho/|DY|)^\alpha$
\be
D_\tau (\rho/|DY|)^\alpha=-\alpha\rho^\alpha|DY|^{-(\alpha+1)}D_\tau |DY|
\ee
and thus, for $\alpha=1/2$
\be
D_\tau (\rho/|DY|)^{1/2}=-\frac{1}{2}\rho^{1/2}|DY|^{-3/2}D_\tau |DY|=-\frac{1}{2}(\rho/|DY|)^{1/2}\d\V,
\ee
where the last equation is obtained using (\ref{dy}). Thus, the evolution equation for
\be
\varphi=(\rho/|DY|)^{1/2}e^{-iY}
\ee
is
\be
i\hbar D_\tau \varphi=-\L\varphi - i\frac{\hbar}{2}{\mathsf{div}} \V\varphi,
\label{eq:all1}
\ee
with the solution
\bea
\varphi(\Y,\tau)&=&\varphi_0(G^{-\tau}(\Y))e^{-\frac{1}{2}\int_0^\tau {\mathsf{div} \V(G^{s}(G^{-\tau}(\Y)))}ds}e^{iS(\tau,\Y)/\hbar}, \nonumber \\
&=&\varphi_0(\X_0)e^{-\frac{1}{2}\int_0^\tau {\mathsf{div} \V(G^{s}(\X_0))}ds}e^{iS(\tau,\Y)/\hbar},
\label{eq:Feyn1}
\eea
Replacing $\tau$ with the classical coordinate time and assuming a flat geometry of spacetime, this solution for the wavefunction reduces to the one derived from the Schr\"odinger equation by Holland (\cite{Holland:2005}, equation 7.3). 
\begin{remark}The equation (\ref{eq:Feyn}) can serve as a template for the path integral formulation of the current theory.
\end{remark}

\subsection{Relationship to Weighted Composition Operators}
Based on (\ref{eq:Feyn}) a group of evolution operators $\W^\tau$ parametrized by proper time can be defined:
\be
\W^\tau \psi=\pi\cdot\psi\circ G^{-\tau}
\ee
where  $\pi:M_x\rar \bC$
\be
\pi(\y)=e^{-\int_0^\tau {\mathsf{div} \V(G^{s}(G^{-\tau}(\Y)))}ds}e^{iS(\tau,\Y)/\hbar}
\ee
The operators $\W^\tau$ belong to the class of the so-called weighted composition operators \cite{sm:1993}.
\section{Special Relativity Case: Dirac Equation}
\label{sec:Dirac}
In this section we show that the scalar wave amplitude $\psi$  of a solution to Dirac equation\footnote{$\psi$ is known to be a solution to the Klein-Gordon equation} satisfies equation  (\ref{eq:all}). Note here that it is only in the divergence part that the equation  (\ref{eq:all}) differs from the probability amplitude equation (\ref{eq:all1}), and we will see below that for Dirac equation the divergence is $0$. We start with the Dirac equation in the form
\be
i\hbar\frac{\p \bm{\psi}}{\p t}=-i\hbar c \alpha_j \cdot \nabla_j \bm{\psi}+\beta mc^2  \bm{\psi}
\label{eq:Dirac}
\ee
where $\bm{\psi}=s\psi$, $s$ is a 4-component spinor, $\psi$ is a scalar function, $\alpha_j$'s and $\beta$ are matrices defined by
\be
\beta=
\begin{bmatrix}
1 & 0 &  0 & 0 \\
0 & 1 &  0 & 0 \\
0 & 0 &  -1 & 0 \\
0 & 0 &  0 & -1 \\
\end{bmatrix}
,
\
\alpha_1=
\begin{bmatrix}
0 & 0 &  0 & 1 \\
0 & 0 &  1 & 0 \\
0 & 1 &  0 & 0 \\
1 & 0 &  0 & 0 \\
\end{bmatrix}
\ee
\be
\alpha_2=
\begin{bmatrix}
0 & 0 &  0 & -i \\
0 & 0 &  i & 0 \\
0 & -i &  0 & 0 \\
i & 0 &  0 & 0 \\
\end{bmatrix}
,
\ 
\alpha_3=
\begin{bmatrix}
0 & 0 &  1 & 0 \\
0 & 0 &  0 & -1 \\
1 & 0 &  0 & 0 \\
0 & -1 &  0 & 0 \\
\end{bmatrix}.
\ee
Let 
\be
n=\sqrt\frac{E+mc^2}{2mc^2}, 
\ee
be the normalization constant. The Dirac spinors for the frame moving with velocity $v$ are \cite{bjorkenanddrell:1965}
\be
u_1=n\begin{bmatrix}
1 \\
0 \\
\frac{p_zc}{E+mc^2} \\
\frac{(p_x+ip_y)c}{E+mc^2}
\end{bmatrix},
u_2=n\begin{bmatrix}
0 \\
1 \\
\frac{(p_x-ip_y)c}{E+mc^2} \\
\frac{-p_zc}{E+mc^2} 
\end{bmatrix},
v_1=n\begin{bmatrix}
\frac{p_zc}{E+mc^2} \\
\frac{(p_x+ip_y)c}{E+mc^2} \\
1 \\
0 \\
\end{bmatrix},
v_2=n\begin{bmatrix}
\frac{(p_x-ip_y)c}{E+mc^2} \\
\frac{-p_zc}{E+mc^2} \\
0 \\
1 
\end{bmatrix}.
\ee
where $E$ is the energy and $p_j$ are components of momentum.
Let 
\bea
\gamma&=&\frac{1}{\sqrt{1-v^2/c^2}}, \nonumber \\
p^2&=&p_z^2+p_x^2+p_y^2.
\eea

Computation yields 
\bea
u_1^c u_1&=&u_2^c u_2=v_1^c v_1=v_2^c v_2=\gamma, \nonumber \\
u_1^c \beta u_1&=&u_2^c \beta u_2=1, \nonumber \\
v_1^c\beta v_1&=& v_2^c \beta v_2=-1, \nonumber \\
u_1^c \alpha_1 u_1&=&v_1^c \alpha_1 v_1=u_2^c\alpha_1 u_2=  v_2^c \alpha_1 v_2=v_x\gamma/c, \nonumber \\
u_1^c \alpha_2 u_1&=&v_1^c \alpha_2 v_1=u_2^c\alpha_2 u_2=  v_2^c \alpha_2 v_2=v_y\gamma/c, \nonumber \\
u_1^c \alpha_3 u_1&=&v_1^c \alpha_3 v_1=u_2^c\alpha_3 u_2=  v_2^c \alpha_3 v_2=v_z\gamma/c.
\label{eq:Diracrel}
\eea
Letting $\bm{\psi}=u_1\psi_u^+$, and premultiplying  (\ref{eq:Dirac}) by $u_1^c$ ,  leads to
\be
i\hbar\frac{\p {\psi_u^+}}{\p t}=-i\hbar  v \cdot \nabla {\psi_u^+}+ \frac{mc^2}{\gamma}  {\psi_u^+}
\label{eq:Dirac1}
\ee

Using proper time
\be 
\tau=\frac{t}{\gamma},
\ee
we obtain
\be
i\hbar D_\tau \psi_u^+=mc^2 {\psi_u^+},
\label{eq:DiracN}
\ee
which is the equation (\ref{eq:all}) for the case of constant velocity with the relativity Lagrangian $\L=-mc^2$.
Calculating similarly, for $\bm{\psi}=u_2\psi_{u}^-,v_2\psi_v^+,v_2\psi_v^-$ we obtain
\bea
i\hbar D_\tau \psi_u^+&=&mc^2 {\psi_u^+} \nonumber \\
i\hbar D_\tau \psi_u^-&=&mc^2 {\psi_u^-} \nonumber \\
i\hbar D_\tau \psi_v^+&=&-mc^2 {\psi_v^+} \nonumber \\
i\hbar D_\tau \psi_v^-&=&-mc^2 {\psi_v^-} 
\eea
where the subscript $u$ denotes the positive energy, $v$ the negative energy solutions, and superscripts $\pm$ refer to spin up and spin down solutions. These resemble equations governing the Dirac particle in rest frame, where $\tau=t$, $v=0$, and reduce to those in the limit.
\begin{remark} This can be interpreted as the fact that when we fix the spin vector $u_k$, the Dirac equation reduces to the equation we derived. It is of interest that the velocity can be interpreted as positive or negative, depending on the sign of the Lagrangian $\pm mc^2$,
just like the Feynman-St\"uckelberg interpretation of positrons moving backwards in time.
\end{remark}
\section{Examples}
\label{sec:exa}

\subsection{  The non-relativistic case of flat 1-dimensional configuration space}
\label{exa:flat1d}
\begin{figure}[h!]
\centering
\includegraphics[clip=true, trim=100 90 200 90,
height=3in, width=5in ]{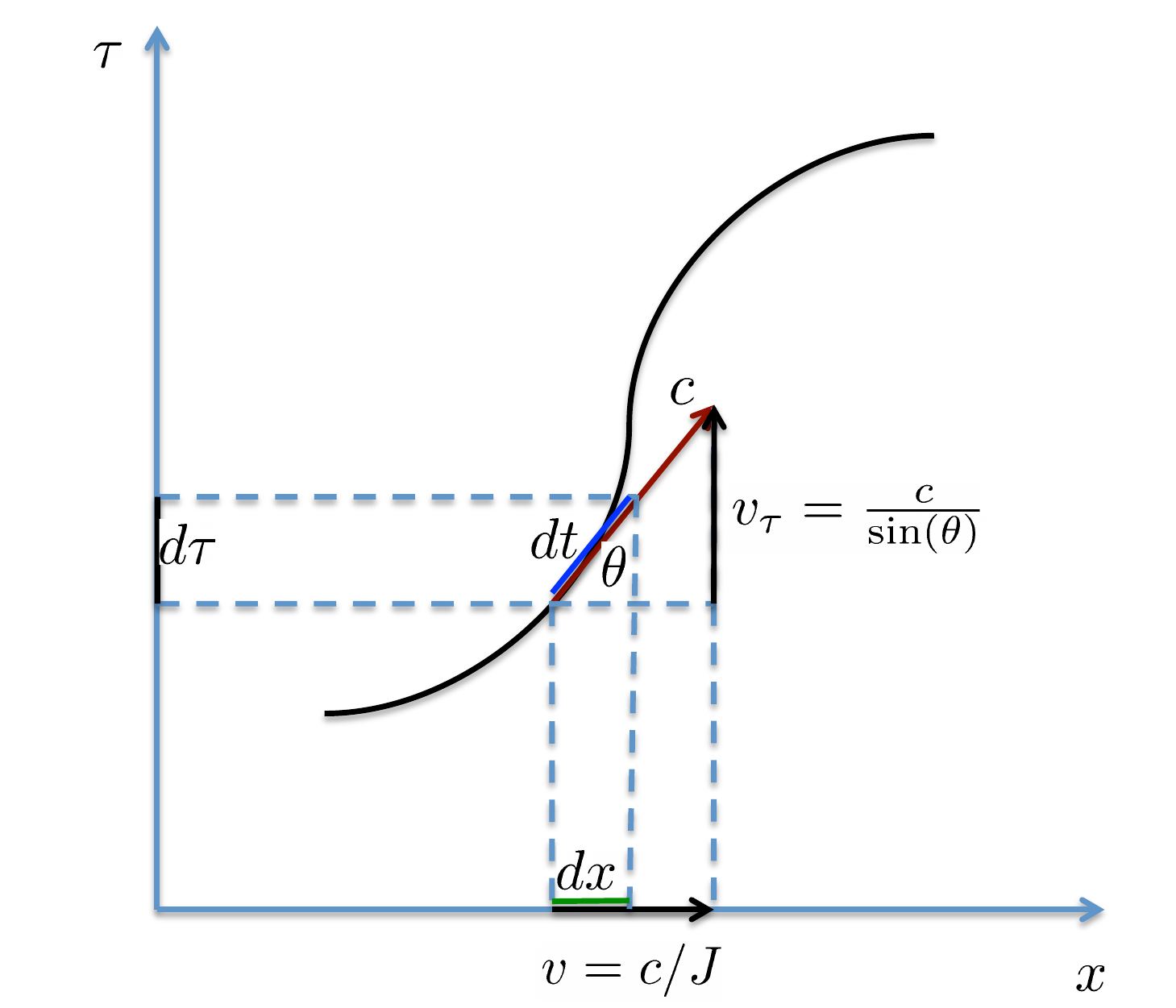}
\caption{The geometry of motion and projection. } \label{fig:TRGeo}
\end{figure}

Consider a 1-dimensional space and proper time, depicted in figure \ref{fig:TRGeo}. The proper time is denoted by $\tau$. 
We denote $v=\dot x=dx/d\tau,$ and assume - for simplicity of notation - that $v$ is positive.
We have
\be
v=\dot x=\frac{c}{J}
\label{eq:vel}
\ee
(see figure \ref{fig:TRGeo}).

 More generally, let $v$ be the norm of the configuration space velocity.
$K$ is exactly the cosine of the angle between the normal to the surface spanned by trajectories in space-proper time and the 
line of sight to the space slice, $1/J=K=v/c$. 

If we set the observable phase $Y$ to satisfy, in any dimension
\be
-\frac{\hbar D_xY}{m}= \vb,
\label{eq:phasevel}
\ee
where $D_xY$ is the reduction of the differential $DY$ to the space slice of $M$, then
\be
\frac{\hbar |D_xY|}{m}= v=\frac{c}{J},
\ee
where $v=|\vb|$, and $m$ is a constant\footnote{In classical optics, $J=c/v$ is the index of refraction. }. Thus the velocity measures the spatial change in the phase of the observable. This corresponds to the non-relativistic case: while the relativistic action with no external potentials given by
\be
S=\alpha\int\sqrt{-c^2\dot t^2+\dot x^2+\dot y^2+\dot z^2}
\ee
where $\alpha=imc$, the classical action is
\be
S_c=\frac{m}{2}\int [\dot x^2+\dot y^2+\dot z^2] dt
\ee
where $t$ is the classical time and $\tau=t$.

Note that with identification 
\be
m=\frac{\rho \hbar}{c},
\ee
 that has dimensions of mass (see Appendix \ref{sec:mass}), we get
\be
\frac{ |DY_x|}{\rho}=\frac{1}{J}\Rightarrow  |DY_x|=\frac{\rho}{J}
\ee
For the wavefunction, we have
\be
 \psi=\frac{\rho}{i|D_xY|e^{iY}}=-iJe^{iY}.
 \label{eq:OW}
 \ee

We consider the flat $1+1$ space-time. As above, we assume that the  wavefunction $\rho$ satisfies
\be
\frac{\p \rho}{\p \tau}+v\frac{\p \rho}{\p x}=0.
\label{eq:rho}
\ee
i.e. $\rho$ is invariant on space-time trajectories.
 We let the observation field $f:\bR^2\rar \bC$ be given by
\be
f=e^{iY}
\ee
The ``observable wavefunction" $\psi$ on the $x$ axis is defined by 
\be
 \psi=\frac{\rho}{i|df/dx|}=\frac{\rho}{i|\frac{\partial Y}{\partial x}|e^{iY}}=\frac{\rho}{iKe^{iY}}.
 \label{eq:OW}
 \ee

We proceed to derive an equation of evolution for $\psi$.
We  set 
\be
D_\tau Y=Y_\tau+vY_x={\tilde{\mathcal L}},
\label{eq:lagr}
\ee
and obtain
\be
\psi_\tau=- v\psi_x -v_x\psi  -i{\tilde\L}\psi,
\label{eq:psit0}
\ee
or, more compactly
\be
i\hbar\frac{\p \psi}{\p \tau}=\H \psi.
\label{QTE1}
\ee
where 
\be
\H=(-i\hbar v\frac{\p}{\p x}+\hbar{\tilde\L})-i\hbar v_x.
\label{eq:H}
\ee 
The equation (\ref{QTE1}), extended to $d$-dimensional configuration space reads 

\be
\psi_\tau=- \vb \nabla \cdot\psi -\nabla \cdot \vb \psi  -i{\tilde\L}\psi,
\label{eq:psit2}
\ee

and has the solution
\be
\psi(\y,\tau)=\psi_0(\x(X^{-\tau}(\y)))e^{-\int_0^\tau {\mathsf{div} \vb(\x(X^{-s}(\y)))}ds}e^{iS(\tau,\y)/\hbar},
\label{eq:Feyn1}
\ee
where $\psi(\z,0)=\psi_0(\z),$ and $X^{-\tau}(\y)$ is the initial position at $\tau=0$ of trajectory landing at $\y$ at $\tau$.
In  next sections we treat the non-relativistic case that makes use of these relationships.
\begin{remark}
\label{rem:comm}
 If we set $Y=-S/\hbar$, the first term on the right side of (\ref{eq:psit0}) is just the quantization of the classical hamiltonian 
 \be
 H=vp+\hbar\tilde\L=vp-\L,
 \ee
  where $p$ gets replaced by $-i\hbar\p /\p x$\
and $\L=-\hbar\tilde\L$ is the lagrangian. 

It is thus clear that in the current theory velocity and momentum are treated separately, like in the context of Dirac equation in Heisenberg representation \cite{barut1984classical}, or Schwinger's variational principle \cite{milton2015quantum}.
\end{remark}

\subsection{The Lagrangian}
\label{sec:lag}
The relativistic lagrangian for a particle with no charge is usually stated as 
\be
\L_0=-\frac{mc^2}{\gamma}.
\ee
In \cite{fock1937proper} Fock developed the so-called proper-time formalism, that utilizes proper time as an independent variable and derived the relativistic Lagrangian for a particle with no charge as 
\be
\L=\frac{m}{2}||\V||^2-\frac{mc^2}{2}
\ee
Since then, the proper time formalism has proved useful in relativistic physics in a variety of contexts \cite{fanchi1993review}. In the examples below, we utilize the Fock Lagrangian  in equation (\ref{eq:all}). For the non-relativistic limit of the harmonic oscillator and particle-in-a-box, we utilize a recent formulation that relates the classical potential $U$ to the metric tensor component $g_{00}$ in mechanics on classical static curved spaces utilizing Gibbons 
formulation \cite{gibbons2015jacobi}:
\be
ds^2=g_{00}c^2dt^2-|d\x|^2.
\ee
Let $U$ be a scalar potential source. As \cite{chanda2018geometrical} shows, the condition $mv^2/2-U<<mc^2$ leads to 
\be
g_{00}\approx 1+\frac{2U}{mc^2}
\ee
\be
\L_c\approx \frac{mv^2}{2}-\frac{mc^2}{2}g_{00}=  \frac{mv^2}{2}-U-\frac{mc^2}{2}
\ee
It is interesting to note that the constant $mc^2/2$ stems from the time-component of the metric tensor. This is of consequence for the zero-point energy calculation in the examples that follow.

\begin{example}[\bf Free Particle] Consider the free particle moving in flat $4$-dimensional space-time with constant 4-velocity $\V$. The divergence $\nabla \cdot \V=0$. Recall that 
\be
t=\frac{\tau}{\sqrt{1-\nu^2/c^2}}=\tau\gamma, \ \gamma=1/\sqrt{1-\nu^2/c^2}.
\ee
 Denote the space components of $\V$ by $\bU$. Since the velocity is constant, the lagrangian reads
\be
\L=-mc^2, 
\ee
and thus from RQTE we get
\be
i\hbar\gamma\frac{\p \psi}{\p t}+i\hbar \bU\cdot \nabla \psi=mc^2\psi,
\label{QTE1free3}
\ee

Let $\u=\bU/\gamma=(U_1/\gamma,U_2/\gamma,U_3/\gamma)$. Then
\be
i\hbar\gamma\frac{\p \psi}{\p t}+i\hbar \gamma \u\cdot \nabla \psi=-{\L}\psi.
\label{QTE1free2}
\ee
\vskip.3cm
\subsection{Dispersion relationship}
\vskip .3cm
We next derive the dispersion relationship for the wave
\be
\psi(\x,\tau)=Ae^{i(\k\cdot \x -\omega t)}.
\ee
From (\ref{QTE1free2}) we get
\be
 \hbar\omega-\hbar\k\cdot\u=\frac{mc^2}{\gamma}=-\L_0. 
\label{disprel}
\ee
Also, with $E=\hbar \omega$ and $\bp=\k\hbar$ we obtain
\be
E=-\L_0+\bp\cdot\u=mc^2\gamma,
\label{reldisp}
\ee
and thus we get the correct relativistic expression for energy. 

\vskip .3cm
\subsection{deBroglie  relationships}
\label{sec:deBrog}
\vskip .3cm
We observe that one of our postulates is conservation of the wavenumber $\rho$ along the spacetime trajectory. Because of the discussion in Appendix \ref{sec:string}, leading to equation (\ref{eq:res}) we assume the natural frequency and wavenumber are related by
\be
 \omega=\rho c \gamma
\ee
which is just the ``coordinate time" version of the relationship (\ref{eq:freq}). The identification $m=\rho \hbar /c$
leads to 
\be
\hbar\omega=mc^2\gamma,
\ee
the first deBroglie  wave-particle relationship. The dispersion equation (\ref{disprel}) now yields
\bea
mc^2\gamma-\hbar\k\cdot\u&=&\frac{mc^2}{\gamma}, \nonumber \\
mc^2-\frac{\hbar\k\cdot\u}{\gamma}&=&\frac{mc^2}{\gamma^2}=mc^2-m\bm\u\cdot \u, \nonumber \\
\frac{\hbar\k\cdot\u}{\gamma}&=&m\bm\u\cdot \u,
\label{disprdeb}
\eea
which in turn gives
\be
\u=\frac{\hbar\k}{m\gamma},
\ee
which is the second deBroglie relationship as we set $\bp=\u m\gamma$.
\vskip .3cm
\subsection{ The relativistic wavepacket}
\vskip .3cm
 The general  solution to (\ref{QTE1free2}) reads

\be
\psi(\x,t)=\psi_{0}(\x-\u t)e^{-i mc^2t/\hbar\gamma}
\label{Gauss1}
\ee
This solution is  physical as long as $\int_\bR |\psi_0| dx$ is finite and thus can be normalized to $1$. 
For simplicity, we restrict to $1$ spatial dimension. Let 
\be
\psi_0(x)=Ae^{-x^2/2\sigma^2}= 2\pi  \sigma A\int_\bR \phi(k)e^{i  kx}dk
\label{trans}
\ee
where 
\be
\phi(k)=e^{-\sigma^2k^2/2}.
\ee

By the second deBroglie relationship derived above, the velocity $u$ is
\be
u=k\hbar/m\gamma
\ee

Integrating over the possible wavenumbers in a wavepacket, we get
\bea
\psi(x,t)&=&2\pi \sigma A\int_\bR \phi(k)e^{ik(x-u t)}e^{-\frac{i}{\hbar} \frac{mc^2 t}{\gamma}}dk \nonumber \\                      
                                     &=&2\pi \sigma A\int_\bR e^{(-\sigma^2k^2/2-it\frac{k^2\hbar}{m\gamma})}e^{-\frac{i}{\hbar} \frac{mc^2 t}{\gamma}}e^{ikx }dk   \nonumber \\
                                       &=&2\pi \sigma A\int_\bR e^{-k^2(\sigma^2/2+i\frac{t\hbar}{m\gamma})}e^{-\frac{i}{\hbar} \frac{mc^2 t}{\gamma}}e^{ikx }dk. \nonumber \\
                                    \nonumber \\
                                    &=&2\pi \sigma A\int_\bR e^{-k^2(\sigma^2+i\frac{2t\hbar}{m\gamma})/2}e^{-\frac{i}{\hbar} \frac{mc^2 t}{\gamma}}e^{ikx }dk. \nonumber \\                              
\label{trans1}
\eea
For wavepacket of small width, where $\gamma(k)\approx const.$, we get
\be
\psi(x,t)
= A\left(\frac{2\pi\sigma}{\sigma^2+i\frac{2t\hbar}{m\gamma}}\right)e^{\frac{-x^2}{\sigma^2+i\frac{2t\hbar}{m\gamma}}}  e^{\frac{i}{\hbar} \frac{mc^2 t}{\gamma}}
\ee
 It is notable that the wavepacket width is suppressed (over the Schr\"odinger wavepacket derived below) due to the
$t\hbar/m\gamma$ term.  In fact, $\sigma^2+i\frac{2t\hbar}{m\gamma}\approx \sigma^2$ as $v\approx c$. Such a suppression was observed in numerical simulations of electrons accelerated  in intense laser fields \cite{maquet2002atoms,su1998relativistic} using the Dirac equation considered on section \ref{sec:Dirac}.

\begin{remark} The solution (\ref{trans1}) can be interpreted in terms of energy as follows:
\bea
\psi(\x,t)&=&A(2\pi \sigma)\int_\bR \phi(k)e^{ik(x-u t)}e^{-\frac{i}{\hbar} \frac{mc^2 t}{\gamma}}dk \nonumber \\
             &=&A(2\pi \sigma)\int_\bR \phi(k)e^{ikx}e^{\frac{it}{\hbar} (-ku \hbar - \frac{mc^2 }{\gamma})}dk \nonumber \\
             &=&A(2\pi \sigma)\int_\bR \phi(k)e^{ikx}e^{-it \frac{E(k)}{\hbar}}dk \nonumber \\
\eea
And we see that the wavepacket is the combination of waves with positive energy. This is in contrast with the Dirac equation, where
the combination of positive and negative energy states is used \cite{huang1952zitterbewegung}.
\end{remark}

\subsection{ Non-relativistic dispersion relationship}
\vskip.3cm
The non-relativistic case is obtained by approximating the Lagrangian with 
\be
\L_0=-\frac{mc^2}{\gamma}+\frac{m u^2}{2}=-\frac{mc^2}{\gamma}+\frac{m k^2\hbar^2}{2m^2 \gamma^2}=-\frac{mc^2}{\gamma}+\frac{k^2\hbar^2}{2m \gamma^2}
\ee
Setting $\gamma \approx 1$, for a single particle in $1$ spatial dimension, we obtain
\be
\psi(x,t)=\psi_{0}(x-u t)e^{i \frac{k^2\hbar t}{2m}}e^{-i mc^2t/\hbar}
\label{Gaussnonr}
\ee
Integrating over the possible velocities in a $1+1$ wavepacket, we get
\bea
\psi(x-u t)&=&a e^{-i mc^2t/\hbar}(2\pi \sigma)\int_\bR e^{- k^2(\sigma^2+i2t\hbar/m)/2}e^{i \frac{k^2\hbar t}{2m}}e^{i k \cdot x }dk  \nonumber \\
&=& ae^{-i mc^2t/\hbar}\left(\frac{\sigma^2}{\sigma^2+it\hbar/m}\right)^{1/2}e^{\frac{-x^2}{2(\sigma^2+it\hbar/m)}},  \nonumber \\
\eea
which is also the result obtained from the Schr\"odinger equation.

The dispersion relationship is
\bea
 \hbar\omega-\hbar ku&=&mc^2-\frac{m u^2}{2} \nonumber \\
 \ \ \ \ \ \ \ \ \ \ \ \ E&=&\hbar \frac{k^2\hbar}{m}-\frac{m k^2\hbar^2}{2m^2}+mc^2\nonumber \\
 &=&\frac{ k^2\hbar^2}{2m}+mc^2
\eea
which, apart from the constant $mc^2$ term is the expression obtained from the Schr\"odinger equation.
\end{example}

\begin{example} \label{ex:harmosc}  We consider the example of the classical one-dimensional harmonic oscillator. The velocity field of the harmonic oscillator in classical space-time ($\tau=t$) is given by (see figure \ref{fig:HarmOsc})
\bea
\dot x&=&A\cos \omega t \nonumber \\
\dot t&\approx& 1.
\eea
\begin{figure}[h!]
\centering
\includegraphics[clip=true, trim=0 0 0 0,
height=3in, width=5in ]{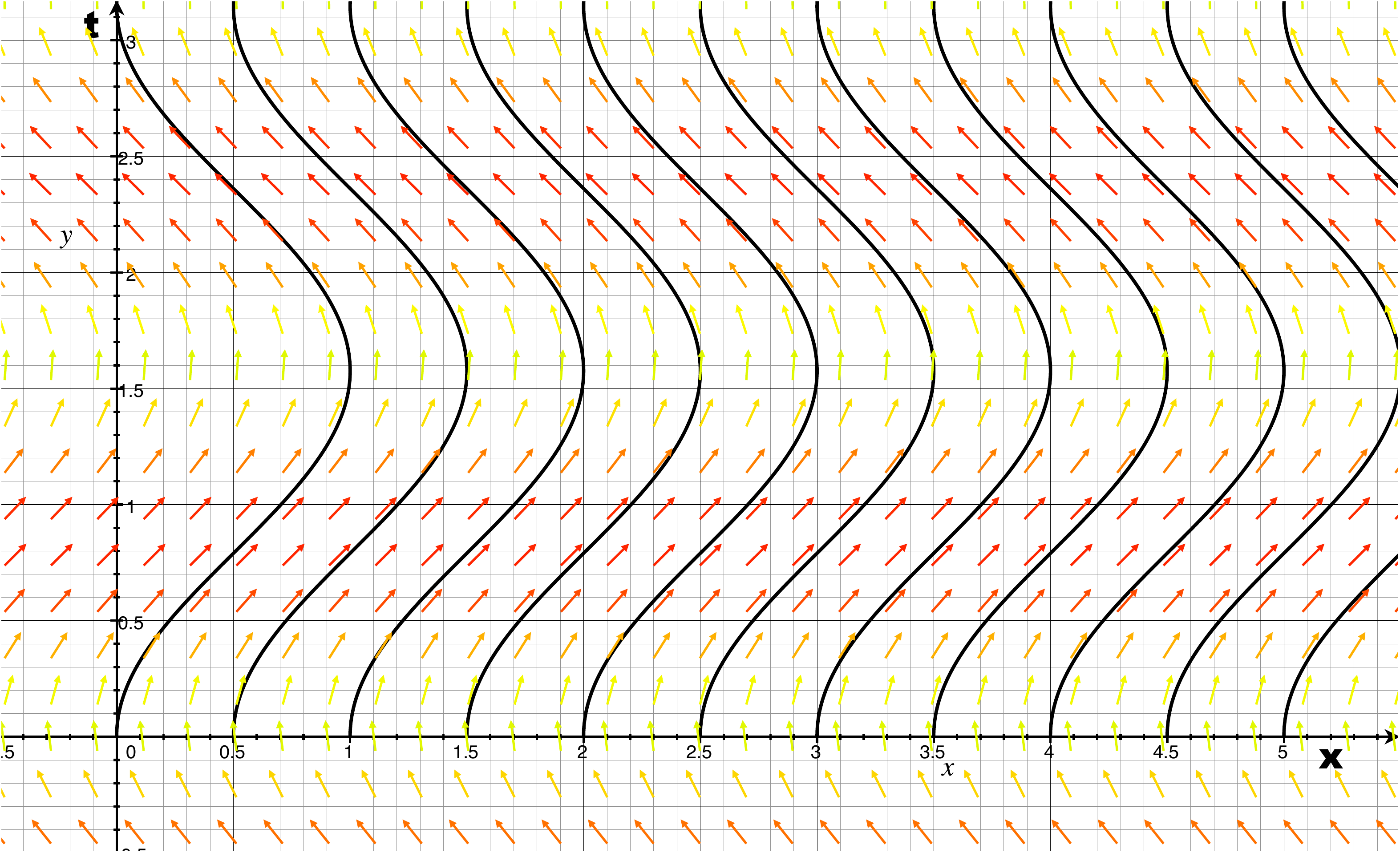}
\caption{The geometry of motion of the harmonic oscillator in classical space-time, with $A=1, \omega=2$. } \label{fig:HarmOsc}
\end{figure} 
Thus, the  divergence of the vector field is $0$.
The Lagrangian reads \cite{chanda2018geometrical}
\be
\L=\L_c-\frac{1}{2}mc^2=\frac{1}{2}mv^2-\frac{1}{2}kx^2-\frac{1}{2}mc^2.
\ee

The trajectory of the harmonic oscillator in space time, taken for simplicity with the initial conditions $(x_0,\dot x_0=0)$  satisfies 
\bea
x(t)&=&x_0\cos \omega t \nonumber \\
\dot x(t)&=& x_0\omega \sin \omega t 
\eea
where $\omega^2=k/m$.
From (\ref{QTE1}), integrating over the period of the trajectory, we have
\be
i\hbar  \int_{\psi_0}^\psi \frac{d\psi}{\psi}= \int_0^T (\lambda+\L) dt.
\label{QTEharc}
\ee
and thus
\be
\psi(T)=\psi(0)e^{-\frac{i}{\hbar}\int_0^T (\lambda+\L)dt}
\ee
In order for $\psi$ to be periodic, we have the condition
\be
\int_0^T (\lambda+\L)dt=n2\pi \hbar
\label{cond}
\ee
where $n\in \bZ$.
Now,
\bea
\int_0^T \L_c dt&=&\int_0^T (\frac{1}{2}mv^2-\frac{1}{2}kx^2) dt \nonumber \\
&=&\int_0^T (\frac{1}{2}m(x_0\omega \sin \omega t)^2-\frac{1}{2}k(x_0\cos \omega t)^2) dt \nonumber \\
&=&\frac{x_0^2}{2}(m\omega^2-k)\int_0^T (\sin \omega t)^2 dt \nonumber \\
&=&0,
\eea
since 
\be
\int_0^T (\sin \omega t)^2 dt=\int_0^T (\cos \omega t)^2 dt,
\ee
and thus, from (\ref{cond})
\be
\lambda T-\frac{1}{2}mc^2T=n2\pi \hbar
\ee
From our previous consideration, using $m=\rho \hbar/c$, we have
\be
\frac{1}{2}mc^2 T=\frac{1}{2}\rho\hbar cT=\frac{1}{2}\omega\hbar T
\ee
where $\omega=\rho c$ is deBroglie wave frequency. Finally, we get
\be
\lambda=\frac{1}{2}\omega\hbar+n\hbar\omega=\omega\hbar(n+\frac{1}{2}).
\ee
which is exactly the standard result on the spectrum of the harmonic oscillator. The zero point energy $\omega \hbar/2$ arises from the oscillation of the observational field, since it comes from the lagrangian term.

Note that the nature of the spectrum is typical of weighted composition operators \cite{gunatillake2007spectrum}, and $n\hbar \omega$ is the spectrum of the underlying composition operator generated by setting $\L=0$.
\end{example}
\begin{example} Consider the example of particle in a box of length ${l}$. Since between impacts with the walls the particle has constant velocity $v$, the classical limit of the Fock lagrangian (ommiting the constant $mc^2$ term) is \cite{chanda2018geometrical}
\be
\L=\frac{1}{2}mv^2
\ee
The particle moves with velocity $v$ between the walls. 
The eigenvalue problem reads
\be
i\hbar v\frac{\p \psi}{\p x}=(\lambda-\L) \psi
\ee
By integrating from $0$ to $l$ 
\be
\frac{\lambda-\L}{\hbar}\frac{l}{v} =n\pi
\ee
i.e.
\be
\lambda-\L=\hbar \frac{n \pi v}{l}
\ee
for $n \neq 0$ as $n=0$ leads to a trivial eigenfunction $0$. De Broglie momentum relationship
\be
p=mv=\frac{h}{\lambda_p}=\frac{2\pi \hbar}{\lambda_p} \Rightarrow v=\frac{2\pi \hbar}{m\lambda_p},
\ee
where $\lambda_p$ is the particle wavelength, leads to 
\be
\L=\frac{1}{2}mv^2=\frac{1}{2}m\frac{4\pi^2 \hbar^2}{m^2\lambda_p^2}=\frac{2\pi^2 \hbar^2}{m\lambda_p^2}
\ee
\be
\lambda=\hbar^2 \frac{n 4\pi^2 }{2m l\lambda_p}-L
\ee
Now we ask for ``spatial" resonance, namely  that the wavelength of string vibration is a subharmonic of the wavelength of the trajectory:
\be
\lambda_p=\frac{2l}{n}
\label{eq:sres}
\ee
we get 
\be
L=\frac{\pi^2 \hbar^2 n^2}{2ml^2}
\ee
\be
\lambda=\frac{\pi^2 \hbar^2 n^2}{m l^2}-L= \frac{\pi^2 \hbar^2 n^2}{2m l^2}.
\ee
Note that the relationship (\ref{eq:sres}) indicates the nonlinearity of the  dynamics: in the case of the harmonic oscillator
treated in example \ref{ex:harmosc}, the frequency of oscillation of the trajectory was matched to the frequency of oscillation of the string. Here, the trajectory motion contains all of the harmonics of the base frequency, and the oscillation of the string can excite any of these.
\end{example}
\section{Discussion and Conclusions}
Starting from several postulates, in this paper we present a relativistic quantum transfer equation (RQTE) governing the evolution of a wavefunction transported by  a 4-velocity field over a spacetime manifold. The key physical assumption is the  existence of a complex scalar field (the horizontal lift) of the dynamics.  When a probabilistic interpretation is sought, the solution of the equation reduces in the non-relativistic limit to the solution of the Schr\"odinger equation. In the special relativity limit, the equation is satisfied by the scalar part of the Dirac spinor. We obtained the classically known spectra from the RQTE formalism in the specific non-relativistic physical cases - the harmonic oscillator and the particle in the box. We additionally considered the problem of the Gaussian wavepacket. The solution of RQTE in this case yields a  prediction that indicates reduction of wavepacket spreading in the limit when velocity approaches the speed of light in vacuum.

Solutions of QRTE lead to evolutions governed by  specific type of transfer operators - the weighted composition operators.
We believe this observation can be useful in further development of the theory and connections between the mathematical literature on such operators (see e.g. \cite{sm:1993}).

It is of interest to note that from our postulates an interesting relationship between RQTE wavefunction and mass (or equivalently energy) emerges. In fact, the concept of mass that arises coincides with the concept of mass stemming from string theory considerations.

Extension of this theory for different spin particles is possible. We hope the developed theory might be useful in numerical methods needed for quantum computing problems.
\appendix
\section{Mass and the Wavefunction}
\label{sec:mass}
 Recall,from our postulates, $\psi=(\rho/i|DY|)e^{-iY}$.  We  note that $\psi$, being a complex number,  is nondimensional.  For $\psi$ to be nondimensional, $\rho$ must have dimensions of a spatial wavenumber, [1/L] where $L$ is the unit of length.

Thus, one can think of the trajectories as carrying waves propagating in time with a certain frequency $\nu$, where $\rho=\nu/c$. 
Since $\rho \hbar/c$ has dimensions of mass, this is to indicate that spacetime trajectories oscillating at higher frequencies
have higher ``mass". Vibrations theory teaches that higher frequencies of oscillation indicate higher stiffness, leading to the idea that mass reflects the ``stiffness" of the underlying trajectory. Interestingly, this is in line with the concept of mass in string theory that is related to tension of the string \cite{tong2009lectures,huggett2015deriving}, see  Appendix \ref{sec:string}.

Our postulates  thus consistent with classical ideas on mass: 1) two objects with a different mass fall at the same speed,
which is the consequence of our assumption that $\rho$ does not affect the velocity of objects in spacetime,
and 2)  it is harder to change the speed of an object  (i.e. bend its trajectory in spacetime) if it has larger mass.
In addition, the relationship $m=\rho \hbar/c$ introduces both $\hbar$ and $c$ into classical mechanics, since the
classical momentum can be written as $p=\rho \hbar v/c$, where $p$ is the linear momentum and $v$ is the velocity of the  particle. 

In fact, the definition of $Y=-S/\hbar$ is necessary precisely to offset the fact that in classical mechanics $m$ and not $\rho$ is used.
The meaning of the constant $\hbar$ emerges as that of a conversion factor between the wavelength of oscillation of a particular space-proper time trajectory, and the 
associated, classically observable, mass. Mass, as defined here, is conceptually the rest mass of special relativity. 

 An analogy offers itself to lend physical intuition about the postulates: the situation is similar to that of observing objects moving at the bottom of a swimming pool through a wavefield on the surface. If the size of the object is much larger than the typical wavelength of the wavefield, their can be seen without an uncertainty proportional to that wavelength - small compared with the size of the object. However, if the object size is comparable to the wavelength, then the uncertainty in observation is large. In our case, the wavelength is $1/\rho=\hbar/ mc$, or precisely  the reduced Compton wavelength.  
\section{Relationship to String Theory}
\label{sec:string}
The ideas in this paper are consistent with deBroglie's wave theory of matter, as we saw in section \ref{sec:deBrog}. But they are also supported by a mechanical model: the existence of the conserved wavenumber $\rho$ indicates that the nature of the underlying object is a string, traveling through space-time at speed $c$. Consider the case of 
\be
S(\tau,\y)=-mc^2\tau
\ee
arising from the relativistic lagrangian $\L=-mc^2$ (see the section \ref{sec:lag} on the lagrangian). The 
frequency of oscillation of the observation field is 
\be
\Omega=\frac{mc^2}{\hbar}
\ee
Since the string has wavenumber $\rho$, the associated natural frequency $\omega_s$ of oscillation of the string is
\be
\omega_s=\rho c.
\label{eq:freq}
\ee
In the case of resonance
\be
1=\frac{\omega_s}{\Omega}=\frac{\rho c \hbar}{mc^2},
\label{eq:res}
\ee
which implies the wavenumber-mass relationship $m=\rho \hbar/c$ that we postulated based on dimensional grounds.
This implies the existence of a matter object of mass $m$ provided there is a resonance between the internal frequency of oscillation of 
the string and the frequency of oscillation of the observation field.

Now we show that this analysis is consistent with the basic ideas in string theory. Consider an open string of length $\mathcal{l}_s$ \cite{Zwiebach:2004}. Let the rest mass per unit length of the string be $\mu_0$. Then the resonance condition reads
\be
mc^2=\mu_0 \l_s c^2=\rho  \hbar c.
\ee
Let $T_0$ be the string tension. From \cite{Zwiebach:2004}, equation (7.26)
\be
\sigma_1T_0=E=mc^2,
\ee
we have
\be
\sigma_1T_0=\rho  \hbar c \Rightarrow T_0=\frac{\rho c \hbar}{\sigma_1 }
\ee
Now we identify 
\be
\rho=\frac{1}{\l_s}.
\ee
This implies that {\sf conservation of $\rho$ on space-time trajectory is equal to the assumption of conservation of open string length}.
We get 
\be
 T_0=\frac{  \hbar c}{\sigma_1 \l_s }
\ee
From (7.64) in \cite{Zwiebach:2004} we have for the string rotational velocity $\omega_s$
\be
\frac{\omega_s}{c}=\frac{\pi}{\sigma_1} \Rightarrow \sigma_1=\frac{\pi c}{\omega_s}=\frac{ \pi\l_s }{2 }
\ee
where the last part emerges from using $\omega_s \l_s/2=c$ from string theory (string ends have speed of light velocity
if it has rotational velocity $\omega_s$), or alternatively by noting that 
\be
\omega_s=2\omega=2\rho c=\frac{2c}{\l_s}
\ee
from the current theory, since in string theory the frequency of repeat of string shape ${\mathbf F}$ for rotating string in spacetime is half the frequency  (twice the period) based on string length.
Thus,
\be
 T_0=\frac{  2\hbar c}{\pi \l_s^2 }
\ee
This is precisely what emerges from  the formula (9.101) in \cite{franklin2010advanced}.

An interesting aspect of this relationship is the nature of the energy term $mc^2$ - this is associated with the observable field oscillation, not with the string oscillation! In contrast,  in string theory, the energy $E$ is the assumed property of the string.

%\section*{References}
\bibliographystyle{unsrt}
\bibliography{schr,KvN}

\begin{thebibliography}{10}

\bibitem{guckenheimer1984nonlinear}
John Guckenheimer and Philip Holmes.
\newblock Nonlinear oscillations, dynamical systems and bifurcations of vector
  fields.
\newblock {\em J. Appl. Mech}, 51(4):947, 1984.

\bibitem{Koopman:1931}
B.O. Koopman.
\newblock Hamiltonian systems and transformation in {H}ilbert space.
\newblock {\em Proceedings of the National Academy of Sciences of the United
  States of America}, 17(5):315, 1931.

\bibitem{Mezic:2005}
Igor Mezi{\'c}.
\newblock Spectral properties of dynamical systems, model reduction and
  decompositions.
\newblock {\em Nonlinear Dynamics}, 41(1-3):309--325, 2005.

\bibitem{Budisicetal:2012}
Marko Budisi{\'c}, Ryan Mohr, and Igor Mezi{\'c}.
\newblock Applied koopmanism.
\newblock {\em Chaos: An Interdisciplinary Journal of Nonlinear Science},
  22(4):047510, 2012.

\bibitem{wilczek:2015}
Frank Wilczek.
\newblock Notes on {K}oopman-von {N}eumann mechanics and a step beyond.
\newblock {\em Unpublished}, 2015.

\bibitem{antoniou2002implementability}
Ioannis Antoniou, Wladyslaw~A Majewski, and Zdzislaw Suchanecki.
\newblock Implementability of {L}iouville evolution, {K}oopman and
  {B}anach-{L}amperti theorems in classical and quantum dynamics.
\newblock {\em Open systems \& information dynamics}, 9(4):301--313, 2002.

\bibitem{gozzi2002minimal}
E.~Gozzi and D.~Mauro.
\newblock Minimal coupling in {K}oopman-von {N}eumann theory.
\newblock {\em Annals of Physics}, 296(2):152--186, 2002.

\bibitem{ghose:2017}
Partha Ghose.
\newblock Continuous quantum-classical transitions and measurement: A relook.
\newblock {\em arXiv preprint arXiv:1705.09149}, 2017.

\bibitem{klein2018koopman}
Ulf Klein.
\newblock From {K}oopman-von {N}eumann theory to quantum theory.
\newblock {\em Quantum Studies: Mathematics and Foundations}, 5(2):219--227,
  2018.

\bibitem{bondaretal:2019}
Denys~I Bondar, Fran{\c{c}}ois Gay-Balmaz, and Cesare Tronci.
\newblock Koopman wavefunctions and classical--quantum correlation dynamics.
\newblock {\em Proceedings of the Royal Society A}, 475(2229):20180879, 2019.

\bibitem{viennot2018schrodinger}
David Viennot and Lucile Aubourg.
\newblock Schr{\"o}dinger--koopman quasienergy states of quantum systems driven
  by classical flow.
\newblock {\em Journal of Physics A: Mathematical and Theoretical},
  51(33):335201, 2018.

\bibitem{joseph2020koopman}
Ilon Joseph.
\newblock {K}oopman-von {N}eumann approach to quantum simulation of nonlinear
  classical dynamics.
\newblock {\em arXiv preprint arXiv:2003.09980}, 2020.

\bibitem{mauro2002koopman}
Danilo Mauro.
\newblock On {K}oopman--von {N}eumann waves.
\newblock {\em International Journal of Modern Physics A}, 17(09):1301--1325,
  2002.

\bibitem{giannakis2019quantum}
Dimitrios Giannakis.
\newblock Quantum dynamics of the classical harmonic oscillator.
\newblock {\em arXiv preprint arXiv:1912.12334}, 2019.

\bibitem{fock1937proper}
V.~Fock.
\newblock Proper time in classical and quantum mechanics.
\newblock {\em Phys. Z. Sowjetunion}, 12:404, 1937.

\bibitem{dirac1996general}
Paul Adrien~Maurice Dirac.
\newblock {\em General theory of relativity}, volume~50.
\newblock Princeton University Press, 1996.

\bibitem{olver2000applications}
Peter~J Olver.
\newblock {\em Applications of Lie groups to differential equations}, volume
  107.
\newblock Springer Science \& Business Media, 2000.

\bibitem{MezicandBanaszuk:2004}
Igor Mezi{\'c} and Andrzej Banaszuk.
\newblock Comparison of systems with complex behavior.
\newblock {\em Physica D: Nonlinear Phenomena}, 197(1):101--133, 2004.

\bibitem{Holland:2005}
Peter Holland.
\newblock Computing the wavefunction from trajectories: particle and wave
  pictures in quantum mechanics and their relation.
\newblock {\em Annals of Physics}, 315(2):505--531, 2005.

\bibitem{sm:1993}
RK~Singh and JS~Manhas.
\newblock {\em Composition operators on function spaces}.
\newblock Number 179. North Holland, 1993.

\bibitem{bjorkenanddrell:1965}
James~D Bjorken and Sidney~D Drell.
\newblock {\em Relativistic quantum mechanics}.
\newblock McGraw-Hill, 1965.

\bibitem{barut1984classical}
AO~Barut and Nino Zanghi.
\newblock Classical model of the dirac electron.
\newblock {\em Physical Review Letters}, 52(23):2009, 1984.

\bibitem{milton2015quantum}
Kimball~A Milton.
\newblock {\em Quantum Action Principle}.
\newblock Springer, 2015.

\bibitem{fanchi1993review}
J.R. Fanchi.
\newblock Review of invariant time formulations of relativistic quantum
  theories.
\newblock {\em Foundations of physics}, 23(3):487--548, 1993.

\bibitem{gibbons2015jacobi}
G.W. Gibbons.
\newblock The {J}acobi metric for timelike geodesics in static spacetimes.
\newblock {\em Classical and Quantum Gravity}, 33(2):025004, 2015.

\bibitem{chanda2018geometrical}
Sumanto Chanda and Partha Guha.
\newblock Geometrical formulation of relativistic mechanics.
\newblock {\em International Journal of Geometric Methods in Modern Physics},
  15(04):1850062, 2018.

\bibitem{maquet2002atoms}
Alfred Maquet and Rainer Grobe.
\newblock Atoms in strong laser fields: challenges in relativistic quantum
  mechanics.
\newblock {\em Journal of Modern Optics}, 49(12):2001--2018, 2002.

\bibitem{su1998relativistic}
Qichang Su, BA~Smetanko, and Rainer Grobe.
\newblock Relativistic suppression of wave packet spreading.
\newblock {\em Optics Express}, 2(7):277--281, 1998.

\bibitem{huang1952zitterbewegung}
Kerson Huang.
\newblock On the zitterbewegung of the dirac electron.
\newblock {\em American Journal of Physics}, 20(8):479--484, 1952.

\bibitem{gunatillake2007spectrum}
Gajath Gunatillake.
\newblock Spectrum of a compact weighted composition operator.
\newblock {\em Proceedings of the American Mathematical Society},
  135(2):461--467, 2007.

\bibitem{tong2009lectures}
David Tong.
\newblock Lectures on string theory.
\newblock {\em arXiv preprint arXiv:0908.0333}, 2009.

\bibitem{huggett2015deriving}
Nick Huggett and Tiziana Vistarini.
\newblock Deriving general relativity from string theory.
\newblock {\em Philosophy of Science}, 82(5):1163--1174, 2015.

\bibitem{Zwiebach:2004}
Barton Zwiebach.
\newblock {\em A first course in string theory}.
\newblock Cambridge university press, 2004.

\bibitem{franklin2010advanced}
Joel Franklin.
\newblock {\em Advanced mechanics and general relativity}.
\newblock Cambridge University Press, 2010.

\end{thebibliography}

\end{document}